\documentclass[usenatbib,usegraphicx,referee]{mn2e}
\usepackage{epsfig}
\def\araa{ARA\&A}
\def\mnras{MNRAS}
\def\apj{ApJ}

\def\aj{AJ}
\def\apjl{ApJL}
\def\aap{A\&A}

\def\baas{BAAS}

\newcommand{\vU}{{\vec{ U}}}
\def\V{{\cal V}}

\begin{document}
\title{Turbulence in the Harassed Galaxy NGC~4254}
\author[Prasun Dutta, Ayesha Begum, Somnath Bharadwaj and Jayaram
  N. Chengalur]{Prasun Dutta$^{1}$\thanks{Email:
    prasun@cts.iitkgp.ernet.in},  
Ayesha Begum$^{2}$\thanks{Email: begum@astro.wisc.edu}, 
Somnath  Bharadwaj$^{1}$\thanks{Email: somnath@phy.iitkgp.ernet.in}
and  Jayaram N. Chengalur$^{3}$\thanks{Email:
  chengalu@ncra.tifr.res.in} 
\\$^{1}$ Department of Physics and Meteorology \&
 Centre for Theoretical Studies, IIT Kharagpur, Pin: 721 302, India,  
\\$^{2}$Department of Astronomy,
University of Wisconsin,
475 N. Charter Street,
Madison, WI 53706 , 
\\$^{3}$ National Centre For Radio Astrophysics, Post Bag 3, 
Ganeshkhind, Pune 411 007, India.} 
\maketitle 

\begin{abstract}
Galaxy harassment is an important mechanism for the morphological
evolution of galaxies  in  clusters. The spiral galaxy NGC~4254 in the  
Virgo cluster is believed to be a harassed galaxy. We have analyzed
the  power spectrum of HI emission fluctuations from NGC~4254 to
investigate  whether it carries any imprint of galaxy harassment.  
The power  spectrum, as determined using the $16$ central  channels
which contain  most of the HI emission,  is found to be 
well fitted by a  power law $P(U)=AU^{\alpha}$  with  $\alpha\ =-\ 1.7\pm
0.2$ at length-scales $1.7 \, {\rm k pc}$ to  $ 8.4 \, {\rm kpc}$. 
This is similar to other normal spiral galaxies which  have a  slope
of $\sim -1.5$ and  is interpreted as arising from two dimensional
turbulence at length-scales larger than the galaxy's
scale-height. NGC~4254 is hence yet another example of a spiral galaxy
that exhibits  scale-invariant density fluctuations out to
length-scales  comparable to the diameter of the HI disk. While a
large variety of possible energy sources like proto-stellar winds,
supernovae, shocks, etc.  have been proposed to produce turbulence, it
is still to be seen whether  these are effective  on length-scales
comparable to that of the entire HI disk. On separately analyzing the
HI power spectrum in different  parts of NGC~4254, we find that the
outer parts have a different slope  ($ \alpha = -2.0\pm0.3$)  compared
to the central part of the galaxy ($\alpha = -1.5\pm0.2$). Such a
change in slope is not seen in other, undisturbed galaxies. We suggest
that, in addition to changing the overall morphology, galaxy
harassment also effects the fine scale structure of the ISM, causing
the power spectrum to have a steeper  slope in the outer parts. 

\end{abstract}


\section{introduction}
\label{sec:intro}
Galaxy harassment (frequent high speed galaxy encounters,
\citealt{MKL96})
 is believed to be an important process in driving the
morphological transformation of spiral galaxies to ellipticals inside
clusters. Typically, the first encounters convert 
a normal spiral galaxy to a disturbed spiral with dramatic
features drawn out from the dynamically cold gas.  The spiral galaxy
NGC~4254, located in the nearby Virgo cluster, is found to have a tail 
\citep{M05} with neutral hydrogen (HI) mass $2.2 \times 10^8 \
M_{\odot}$ within a 
distance of $120 \ {\rm kpc}$ from the galaxy. This gaseous tail,
without any  stellar counterpart, is believed to have been
produced by an act of galaxy harassment \citep{HGK07}. Each act of
harassment  has the potential to induce a burst of star formation and
to  change the internal properties of the galaxy, including the 
properties of the inter stellar medium (ISM).  Here we study the effect
of the harrasment on the fine scale structure of the ISM. We use the
power spectrum of HI intensity fluctuations to quantify the fine
scale structure.

Power spectrum analysis of the HI $21$-cm intensity fluctuation has
been  widely used to probe the ISM
\citep{CD83,GR93,SBH98,SSD99,DD00,EK01,ES04I, ES04II}. These studies,
mainly of our Galaxy and its satellites,   find a  power law  power
spectrum $P(U)=A \ U^{\alpha}$ with index $\alpha$ in the range $-2.5$
to $-3.0$.  This scale invariant power spectrum is interpreted as
arising from  turbulence in the ISM. 

Recently \citet{AJS06} have presented a visibility based formalism for
determining the power spectrum of  galaxies with
extremely weak HI emission. This has been applied to a sample of  dwarf
and   spiral galaxies (\citealt{DBBC08}; \citealt{DBBC09a};
\citealt{DBBC09b}). These studies indicate a dichotomy in 
$\alpha$ with values $\sim -1.5$ in some galaxies and $\sim -2.5$ in
others.  The two values $\sim -1.5$ and $\sim -2.5$ are interpreted as
arising respectively from two (2D) and three (3D) dimensional
turbulence.  Some support for this interpretation has been provided
by analysis of the fluctuation power spectrum for NGC~1058 where a
transition from 2D to 3D turbulence is observed at an angular
scale corresponding to the scale height of the galaxy's disk.

NGC~4254 the galaxy whose power spectrum we estimate in this paper,
is a  lopsided spiral galaxy (morphological type SA(s)c), with  an 
inclination of $\sim 42^{\circ}$ \citep{PH93}. The galaxy is  
located  at a distance of 1 Mpc from the core of Virgo cluster and
is  believed to be falling into the cluster with a relative velocity 
of $1300 \ {\rm km\ s}^{-1}$ \citep{VH05}. The distance to this 
galaxy is estimated to be 16.7 Mpc \citep{M07}; at this distance
1$^{\prime\prime}$ corresponds to 81 pc. 
\begin{figure}
\begin{center}
\includegraphics[angle=0,width=2.8in]{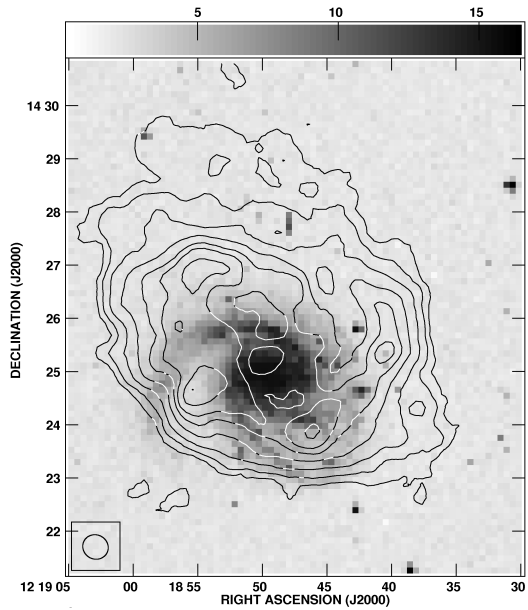}
\end{center}
\caption{
The $6.5^{\prime\prime} \times 8.0^{\prime \prime}$ resolution
integrated HI column  density map of NGC~4254 (contours) is overlayed
with the optical image of the galaxy (density).  The contours levels 
are  3., 5., 10., 20., 25., 35., 40. and 45. $\times
10^{20}$ atoms cm$^{-2}$. }
\label{fig:mom0}
\end{figure}

\section{Data and Analysis}
\label{sec:data}

We have used  archival HI data  of  NGC~4254  from the Very Large
Array (VLA). The   observations had been carried out   on
$4^{th}-5^{th}$ March 1992 using the  C configurations of the VLA
\citep{PH93}.  
The data was downloaded from the VLA archive and
reduced in the usual way using  standard tasks in classic
{\small AIPS}\footnote{NRAO Astrophysical Image Processing System,  
a commonly used software for radio data processing.}. 
The HI emission from NGC~4254 spans over 30   channels i.e,
($18$ to $47$) from $1401.7$ MHz ($2253$ km s$^{-1}$) to $1416.4$ MHz
($2562$ km s$^{-1}$) of the $63$ channel spectral cube.  A frequency
width of $48.8$ k Hz for    each channel in the data cube corresponds
to  the velocity resolution of $10.32~ \, {\rm   km~s}^{-1}$.     The
continuum from the galaxy was subtracted from the data in  the
{\it{uv}} plane using the {\small AIPS} task {\small  UVSUB}. The
resulting continuum subtracted data was used for the 
subsequent analysis. Figure~\ref{fig:mom0} shows a  integrated HI
column density (Moment 0) map of NGC~4254 made from this data.  
The angular extent of the HI distribution in Figure~\ref{fig:mom0} 
is measured to be $6.5' \times 8.0'$  at a  column density of $10^{19}
\, {\rm atoms \, cm}^{-2}$, which is comparable to it's
optical diameter \citep{DV91}.

\begin{figure}
\begin{center}
\includegraphics[angle=-90,width=2.8in]{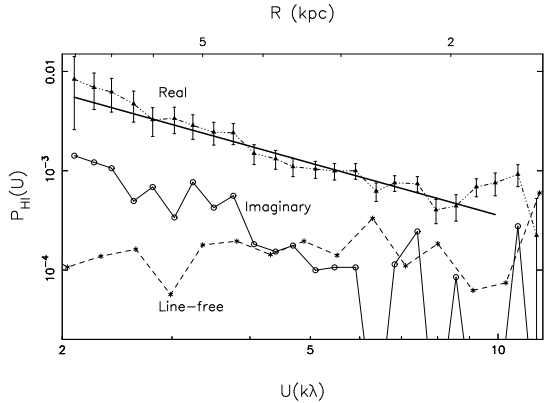}
\caption{Real and imaginary parts of the
 observed value of the HI   power spectrum estimator for channels
 27-42 with $N=16$.
  $1 \sigma$ error-bars are  shown only   for the real part. The real
 part of the estimator from line-free channels is also shown. }
\end{center}
\label{fig:powsp24}
\end{figure}

\citet{AJS06} and \citet{DBBC09a} contains a detailed discussion of 
the visibility based HI power spectrum estimator $\hat{\rm P}_{\rm
  HI}(\vU) =
\langle\ \V_{\nu}(\vU)\V^{*}_{\nu}(\vU+\Delta \vU)\ \rangle$, hence we
present only a brief discussion here.   Here $\vU$ refers to a
baseline, i.e. the antenna  separation projected  in the plane
perpendicular to the direction of   observation,  measured in units of
the observing  wavelength   $\lambda$. It is common practice to
express the dimensionless    quantity $\vU$ in units of kilo
wavelength  (k$\lambda$). Every visibility $\V_{\nu}(\vU)$   is
correlated with all other visibilities  $\V^{*}_{\nu}(\vU+\Delta \vU)$
within a disk $|\Delta \vU| < (\pi  \theta_{0})^{-1}$, where
$\theta_0$ is the angular extent of the  galaxy.   
The correlations are averaged over different   $\vU$
directions assuming that  the signal is statistically isotropic in the
plane of the galaxy's image.
To increase the signal to noise ratio we further average the
correlation in  bins of $U$ and over all  frequency channels with  HI
emission.  The expectation value of the estimator $\hat{\rm P}_{\rm
  HI}(\vU)$ is real,  and it is 
the convolution of the HI  power spectrum $P_{\rm HI}(U)$ with  a window 
function  $|\tilde{W}_{\nu}(U)|^{2}$.  The window function, which  
quantifies the effect of the large-scale HI distribution and the
finite angular extent of  the galaxy, is peaked around $U=0$ and has a 
width of order $(\pi\theta_{0})^{-1}$ beyond which
$|\tilde{W}_{\nu}(U)|^{2}\sim 0$. Analytic considerations and
simulations  \citep{DBBC09a} show that beyond a baseline $U_m= 3.5
\theta_0^{-1}$ the convolution does not effect
the shape of the HI power spectrum, and we may directly interpret  the
real part of the estimator $\hat{\rm P}_{\rm  HI}(\vU)$ as the HI
power spectrum $P_{\rm HI}(U)$.

The estimator $\hat{\rm P}_{\rm  HI}(\vU)$ also has a small imaginary
component  arising mainly from noise. 
The $1-\sigma$ error-bars for the estimated power spectrum is a sum,
in quadrature,  of 
contributions from two sources of uncertainty. At small $U$ the
uncertainty is dominated by the sample variance which comes from the
fact that we have a finite  and limited  
number of independent estimates of the true power spectrum. At
large $U$, it is dominated by the system noise in each visibility. 

\section{Results and Discussion}
\label{sec:discuss}
 Figure~\ref{fig:powsp24} shows the real and imaginary parts of
 $\hat{\rm P}_{\rm  HI}(\vU)$ evaluated using $16$ 
 channels from $27$ to $42$ which have relatively high HI emission.
Here the estimator  $\hat{\rm P}_{\rm  HI}(\vU)$ was separately evaluated 
for each channel and then averaged over all $16$ channels to increase
 the signal to noise ratio. 
 As
expected  from the theoretical considerations (\citealt{AJS06}), the
imaginary part is well suppressed compared to  the real part.  We also 
estimate $\hat{\rm P}_{\rm  HI}(\vU)$ using  the line free  channels 
to test  for  any  contribution from the residual continuum.
This is found to be much smaller than the signal
(Figure~\ref{fig:powsp24}) indicating that the continuum has been
adequately 
 subtracted out. 

The power law $P_{\rm HI}(U)=AU^{\alpha}$, with  $\alpha\ =-\ 1.7\pm
0.2$  is found to 
give a good fit to the  HI power  spectrum for the $U$ range 
$2.0 \, {\rm k}  \lambda$ to $10.0 \, {\rm k}  
\lambda$ (Table~\ref{table:t1}).  The best-fit power-law 
and the $1-\sigma$ error-bar on $\alpha$ 
were determined using   $\chi^2$ minimization  \citep{AJS06,DBBC09b}. 
We use $D=16.7 \ {\rm Mpc}$ (distance to the   galaxy)  to convert a
baseline $U$ to a length-scale $D/U$ in the plane of the galaxy's
image.  The $U$ range  $2.0 \, {\rm k}  \lambda$ to $10.0 \, {\rm k}   
\lambda$ corresponds to  the range of length-scales  $8.4$ to $1.7 \
{\rm kpc}$.  \citet{DBBC09a} have broadly classified the observed
turbulence   in their galaxy sample   as 2D turbulence at
  the large scales in the plane of the galaxies disk and 3D turbulence
  at scales smaller than the scale height of the disk. 
They also show that the slope of the power spectrum changes from $\sim
-1.5$ to $\sim -2.5$ as one goes from 2D to 3D turbulence. For the present
  case the largest  length-scale   ($8.4 \, {\rm kpc}$) 
is  definitely larger than the  typical HI scale heights within the 
Milky-Way \citep{LH84,WB90} and in  external spiral galaxies
(e.g. \citealt{narayan}). It is thus quite reasonable to 
conclude that   the slope  $\alpha = -1.7 \pm 0.2$ is  of 2D 
turbulence  in the plane of the galaxy's disk. The fact that we do not
observe the transition to 3D turbulence, which is expected   to
occur at a baseline $U=D/\pi  z_h$ \citep{DBBC09a},   allows us to
place an upper limit on the galaxy's scale height $z_h 
\le 2.6\ {\rm  k pc}$. 

Our present observation is another confirmation (Dutta et al 2008,
2009a 2009b) of the fact that spiral galaxies exhibit 
scale-invariant density fluctuations that extend to length-scales 
of  $\sim 10 \, {\rm kpc}$  (eg. NGC~628, NGC~1058) which is
comparable to the diameter of the HI disk. While a large variety of
possible energy sources like proto-stellar winds, supernovae, shocks, etc. 
have been proposed to drive turbulence  \citep{ES04I},  it is still to
be seen whether these are effective  on length-scales as  large
as $10 \, {\rm kpc}$.

\begin{figure}
\begin{center}
\includegraphics[angle=0,width=6in]{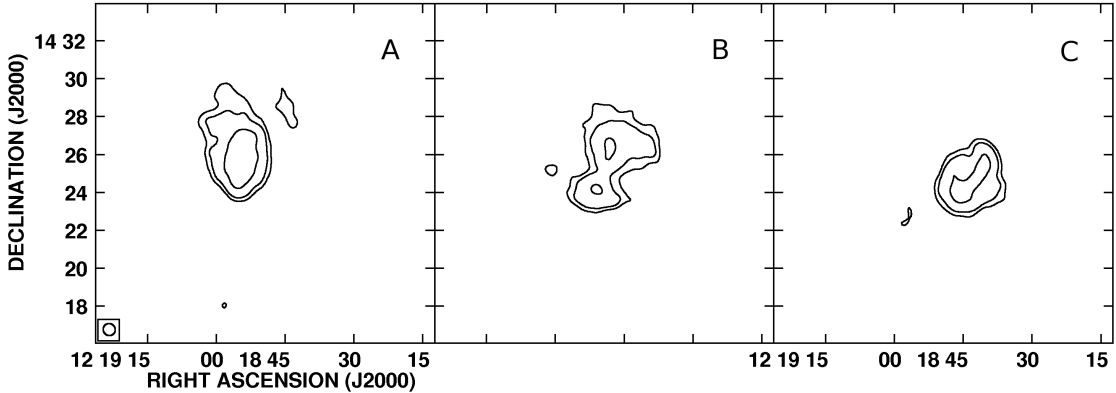}
\caption{Integrated HI column density maps of the galaxy NGC~4254 using
  data cubes A, B and          
   C. Note the diagonal movement of the centroid of emission from North
  east (A) to South west (C). The contours levels are  5., 8. and
  12. $\times 10^{20}$ atoms cm$^{-2}$.} 
\label{fig:abc}
\end{center}
\end{figure}

\begin{figure}
\begin{center}
\includegraphics[angle=-90,width=2.8in]{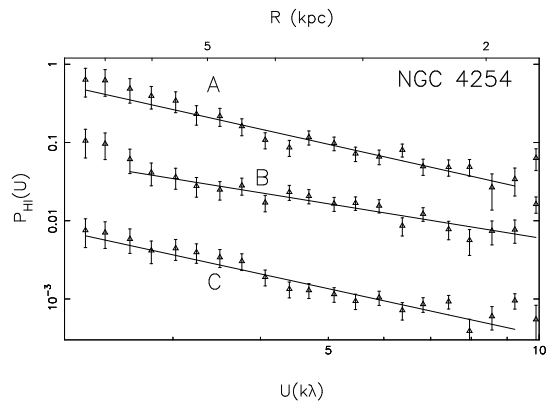}
\caption{HI power spectrum for data cubes A, B and C plotted with
  arbitrary offsets to prevent them from overlapping. }
\label{fig:aps}
\end{center}
\end{figure}

\begin{figure}
\begin{center}
\includegraphics[angle=0,width=6in]{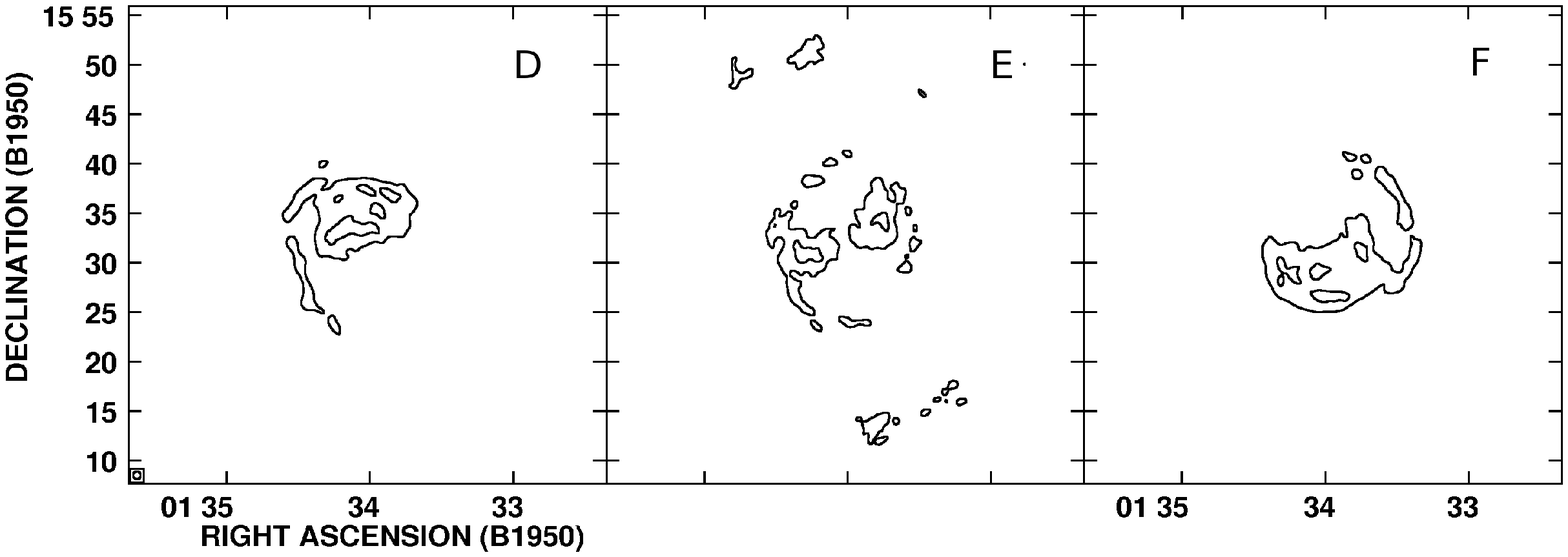}
\caption{Integrated HI column density maps of the galaxy NGC~628 using
  data cubes D, E and  
  F. Note the diagonal movement of the centroid of emission from North
  east (D) to South west (F). The contours levels 
are  6.,  12. and 18. $\times 10^{20}$ atoms cm$^{-2}$. } 
\label{fig:DEFM}
\end{center}
\end{figure}

\begin{figure}
\begin{center}
\includegraphics[angle=-90,width=2.8in]{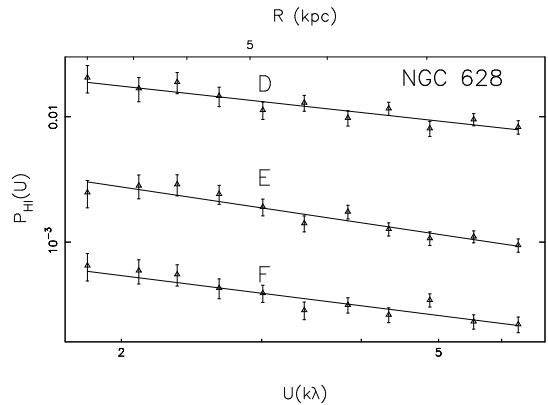}
\caption{Power spectra of the HI emission for the galaxy NGC~628 is
  shown for D (channels 108-119), E (channels 120-131) and F (channels
  132-143) with  arbitrary offsets to prevent them from
  overlapping. Details of the 
  data used for this estimation can be found in \citet{DBBC08}}
\label{fig:628PS}
\end{center}
\end{figure}

Galaxy harassment is expected  to have  different effects on the
inner and outer parts of the galaxy. While gas is stripped 
from the outer parts, the inner part looses angular momentum and
gradually collapses to the center through repeated galaxy encounters. 
We can selectively study different parts of NGC~4254, whose rotation
axis is tilted at  $42^{\circ}$  to the line of sight, by considering
different velocity channels. Our analysis till now has used only the
central $16$ channels, we now use the central $24$ channels ($23-46$)
for the subsequent analysis.  We  construct  
$3$  different data cubes namely A, B and C containing channels $23-30$,
$31-38$ and $39-46$ respectively. We can now separately probe the 
North east,  central and South west parts of the galaxy
(Figure~\ref{fig:abc}) using these three data cubes. For each data
cube,  we  evaluate the HI power spectrum using individual channels
and then average over the channels in the respective cubes. We estimated 
$U_m$ seperately for A, B and C and the best fit
power law is obtained for $U \ge U_m$ only. We find
that for each  data cubes the HI power spectrum is well fit by a
power law (Figure~\ref{fig:aps}), the details being shown in  
Table~\ref{table:t1}. We find that the slope $\alpha$ is 
$-2.0\pm 0.3$ for A and C which probe the outer parts of the disk
while it is  $-1.5\pm 0.2$ in B which probes the central region. 
To verify  that this change in slope is due to harrassment, 
we also consider the HI power spectrum of the  spiral galaies NGC~628
and NGC~1058 \citep{DBBC08,DBBC09b} which are not undergoing galaxy
harassment.  Figure \ref{fig:DEFM} shows  the regions  of the spiral
galaxy    NGC~628  corresponding to  each of the channel range
108-119 (D),  120-131 (E) and  132-143 (F).  We find that the 
power spectra of these three data cubes (Figure~\ref{fig:628PS})
all  have the same slope $\sim -1.6$,  which is also similar to the
slope of the central part of NGC~4254. The results are similar for 
NGC~1058 and hence we do not explicitly show these here. 
 Based on this we conclude
that the difference in slope 
between the inner and outer parts of NGC~4254 is a consequence of
galaxy harassment.  Not only does galaxy harassment affect the
global morphology of the galaxy, it also affects the fine scale
structure in the ISM as reflected by  HI power spectrum.

We note that NGC~4254 have an inclination of $42^{\circ}$, on the other
hand NGC 628 and NGC~1058 are more face-on galaxies (inclination
$\sim 10^{\circ}$). In our analysis we selectively study different
parts of a galaxy by considering different velocity channel
ranges. However, since face-on galaxies do not have much
range in radial velocity from the rotation curve, the spatial extent
of HI in NGC~628 and NGC~1058 will not be very different for the three
velocity channel ranges, which is unlike the case for NGC~4254.
Hence it is likely that we failed to find a change of slope in NGC~628
and NGC~1058 because of this effect of inclination.
However, as seen in Figure~\ref{fig:DEFM} for NGC~628, all three data
cubes D, E and F have a significant contribution of HI from the outer
region of NGC~628, leading us to believe that the power spectrum
has a slope of $\sim−1.6$ in the outer parts of NGC~628. This is similar
to the value of slope seen in the central parts of NGC~4254 and
significantly different from the slope in the
outer parts of NGC~4254. Hence it indicates an impact of harassment in
NGC~4254. Further analysis of spiral galaxies with large inclination
angles would possibly be able to resolve this issue.


We currently do not have  an understanding of how galaxy harassment
caused a steepening of the HI power spectrum in the outer parts of the
galaxy.  Theoretical modeling and the analysis of other Virgo cluster 
spiral galaxies are  needed for further progress in this direction. 

\begin{table}
\centering
\begin{tabular}{|c|c|c|c|c|c|}
\hline 
Data & Channels & $U_{m}$ & $U$ range & $\alpha$   \\
& & (k $\lambda$) & (k $\lambda$) & \\
\hline
\hline
16 central  & $27-42$ & $1.8$ & $2.5 -10.0$ & $-1.7\pm 0.2$ \\
channels &&&& \\
\hline
A & $23-30$ & $1.3$ & $2.0 -10.0$ & $-2.0\pm 0.3$ \\
B & $31-38$ & $2.2$ & $2.5 -10.0$ & $-1.5\pm 0.2$ \\
C & $39-46$ & $1.3$ & $2.0 -10.0$ & $-2.0\pm 0.3$ \\
\hline
 \hline 
\end{tabular}
\caption{Results of the power spectrum analysis of NGC~4254 for
  different data cubes spanning over different velocity ranges.}  
\label{table:t1}
\end{table}

\section*{Acknowledgments}

P.D. is thank full to Sk. Saiyad Ali, Kanan Datta, Tapomoy Guha
Sarkar, Tatan Ghosh, Suman Majumder, Abhik Ghosh Subhasis Panda and
Prakash Sarkar for use full discussions. P.D. would like to
acknowledge HRDG CSIR for providing financial
support. S.B. would like to acknowledge financial support from BRNS,
DAE through the project 2007/37/11/BRNS/357. The data presented in
this paper were obtained from the National Radio Astronomy Observatory
(NRAO) data archive. The NRAO  is a facility of the US National
Science Foundation operated under cooperative agreement by Associated
Universities, Inc.

\end{document}